\documentclass[twocolumn,english]{revtex4-1}
\usepackage[utf8x]{inputenc}
\usepackage{textcomp}
\usepackage{graphicx}
\usepackage{hyperref}
\usepackage{amssymb}
\usepackage{latexsym}
\usepackage{babel}

\usepackage{fancyhdr}
\fancypagestyle{firstpage}{%
  \lhead{This is the version of the article before peer review or editing, as submitted by an author to {Jpn. J. Appl. Phys.} IOP Publishing Ltd is not responsible for any errors or omissions in this version of the manuscript or any version derived from it. The Version of Record is available online at \href{https://doi.org/10.7567/JJAP.52.08JE14}{https://doi.org/10.7567/JJAP.52.08JE14}}
  \rhead{}
}

\begin{document}
\title{Nature of V-Shaped Defects in GaN}

\author{Vladislav Voronenkov$^{1,2}$ } \email{voronenkov@mail.ioffe.ru}

\author{Natalia Bochkareva$^{2}$ }

\author{Ruslan Gorbunov$^{2}$}

\author{Philippe Latyshev$^{3}$}
\author{Yuri Lelikov$^{2}$}
\author{Yuri Rebane$^{2}$}
\author{Alexander Tsyuk$^{2}$}
\author{Andrey Zubrilov$^{2}$}

\author{Yuri Shreter$^{2}$} \email{y.shreter@mail.ioffe.ru}

\affiliation{$^{1}$St. Petersburg State Polytechnical University, Polytechnicheskaya 29, St. Petersburg
194251, Russia \\
$^{2}$Ioffe Physical Technical Institute, Polytechnicheskaya 26, St. Petersburg 194021, Russia \\
$^{3}$Fock Institute of Physics, St. Petersburg State University, St. Petersburg 198904,
Petrodvorets, Ulyanovskaya 1, Russia}

\begin{abstract}
GaN films with thicknesses up to 3 mm were grown
in two custom-made halide vapor phase epitaxy (HVPE)
reactors. V-shaped defects (pits)
with densities from 1~cm$^{-2}$ to 100~cm$^{-2}$ were found on the surfaces of the films.
Origins of pit formation and the process of pit overgrowth
were studied by analyzing the kinematics of pit evolution.
Two mechanisms of pit overgrowth were observed.
Pits can be overgrown intentionally by
varying growth parameters to increase the growth rate of pit facets.
Pits can overgrow spontaneously if a fast-growing facet nucleates
at their bottom under constant growth conditions.
\end{abstract}


\maketitle
    \thispagestyle{firstpage}

\section{Introduction} 

The lack of widely available native substrates is an important problem of the GaN industry.
Halide vapor phase epitaxy (HVPE) is a common method for producing thick GaN films,
suitable for making GaN substrates. 
A typical problem in HVPE growth is the formation of V-shaped defects or pits on the film surface during the growth.
Their pits appear to be inverted pyramids or cones.
Their size  varies from nanometers \cite{chen1998pit} to several millimeters in thick layers~\cite{richter2012pits}.
Generally, pits should not be formed, as they reduce the effective area of the substrate.
However, intentionally produced pits can be used as a dislocation sink to reduce dislocation density \cite{motoki2007dislocation}.
In this work, the GaN films grown in a custom-made HVPE reactor were studied in order to understand why do pits form,
how do growth parameters affect the pit shape and how do pits overgrow.

\section{Experimental Procedure}

GaN films were grown in two custom-made vertical HVPE reactors with resistive heating.
The   multi-wafer 6x2'' reactor was  optimized for films with thicknesses from 10~$\rm \mu m$ to 400~$\rm \mu m$.
The other reactor optimized for long growth processes was employed to grow GaN boules with a millimeter range thickness and a diameter of 3''.

The operating conditions used were similar in both reactors.
Sapphire substrates with $\langle0001\rangle$ orientation were used.
The substrate surface was nitridated in an ammonia atmosphere at 1060~$^{\circ}$C.
Then, a low-temperature buffer GaN layer was deposited at 600~$^{\circ}$C and a total pressure of 250~Torr.
The thickness of this layer was about 1 $\rm \mu m$.
Subsequently, the main growth process was performed at a reactor pressure of 800~Torr in a temperature range of 990~$^{\circ}$C -- 1150~$^{\circ}$C.
Nitrogen was employed as a carrier gas. The source gases used were GaCl and NH$_3$.
To grow thick crack-free films with a smooth surface, the two-stage growth process \cite{tavernier2002progress,tsyuk2011effect} was applied.

 Optically polished cross-sections of the films were prepared to observe the pits.
Samples were cut using a diamond disk saw.
The cut was made as near as possible to the investigated pits.
Then, the samples were lapped and polished to make the surface of the cut optically smooth.
Diamond grits with sizes of 60~$\rm \mu m$  and 7~$\rm \mu m$ were used for coarse and fine lapping processes, respectively.
Polishing was conducted on a felt pad with a 1~$\rm  \mu m$ diamond grit.

\section{Pit Formation and Evolution}

\begin{figure}[!t]
\centering
\includegraphics[width=\linewidth]{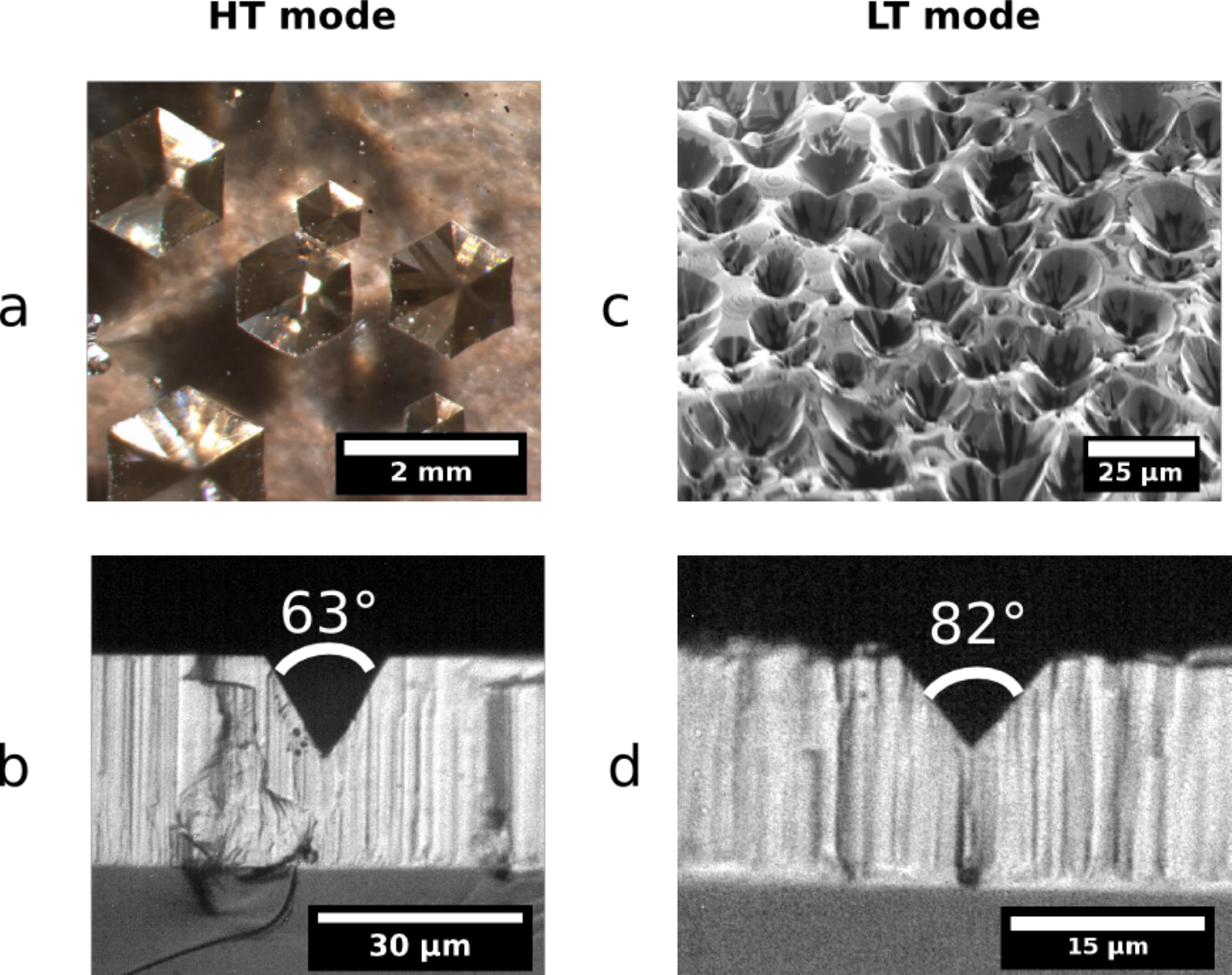}
\caption{Surface morphologies of GaN films grown in High Temperature (HT) and Low Temperature (LT) modes:
(a)~Micrograph of a film grown in HT mode.
(b)~Cleavage of a HT film crossing a pit.
(c)~Focused ion beam image of a film grown in LT mode.
(d)~Cleavage of a LT film crossing a pit.
}
\label{modes}
\end{figure}

A series of 50 films with thicknesses from 30~$\rm \mu m$ to 100~$\rm \mu m$ were grown at different temperatures and growth rates.
Two growth modes with different surface morphologies were observed.
These modes will be further referred to as  high temperature (HT) and  low temperature (LT) modes.
The detailed descriptions of these growth modes and the transition between them is given in \cite{voronenkov2013modes}.

The films grown in the HT mode had a smooth surface (Fig.~\ref{modes}(a)).
 The  average roughness measured by AFM was 20~nm over the area of  $50\times50\,\rm \mu m^2$.
A network of buried cracks could be observed inside such films, indicating  relatively high growth stress.

The average density of pits was about 1 $\rm  cm^{-2}$.
 Pits were hexagonal, with an opening angle of 63 degrees that corresponds to the $\langle11\overline{2}2\rangle$ plane (Fig.~\ref{modes}(b)).

The films grown in the LT mode had a rough surface covered by pits  and hillocks (Fig.~\ref{modes}(c)).
The opening angle of pits was in the range of 80-100 degrees (Fig.~\ref{modes}(d)).
The pits were conical and were composed of $\langle1\overline{1}02\rangle$
and  $\langle11\overline{2}3\rangle$
 facets.
The pits were distributed evenly and occupied a large part of the film surface.
The films were typically crack free.

Combining LT and HT growth modes in a two-stage process allowed us to grow crack-free films with thicknesses up to 3~mm.
The growth was performed in the 1x3'' reactor.
The first stage was performed in the LT mode to promote  low growth stress.
The thickness of the material deposited during the first stage ranged from 50~$\rm \mu m$ to 200~$\rm \mu m$.
The second stage was performed in the HT mode. 
During the second stage, the bulk material was grown. 
More than 30 GaN films with thicknesses  from 2~mm to 3~mm were grown.
All the films had pits with densities from 1~$\rm cm^{-2}$  to 100~$\rm cm^{-2}$,
limiting their application as a substrate.
To find when and why these pits were formed, the history of the pits was investigated in every particular film.

\begin{figure}
   \includegraphics[width=\linewidth]{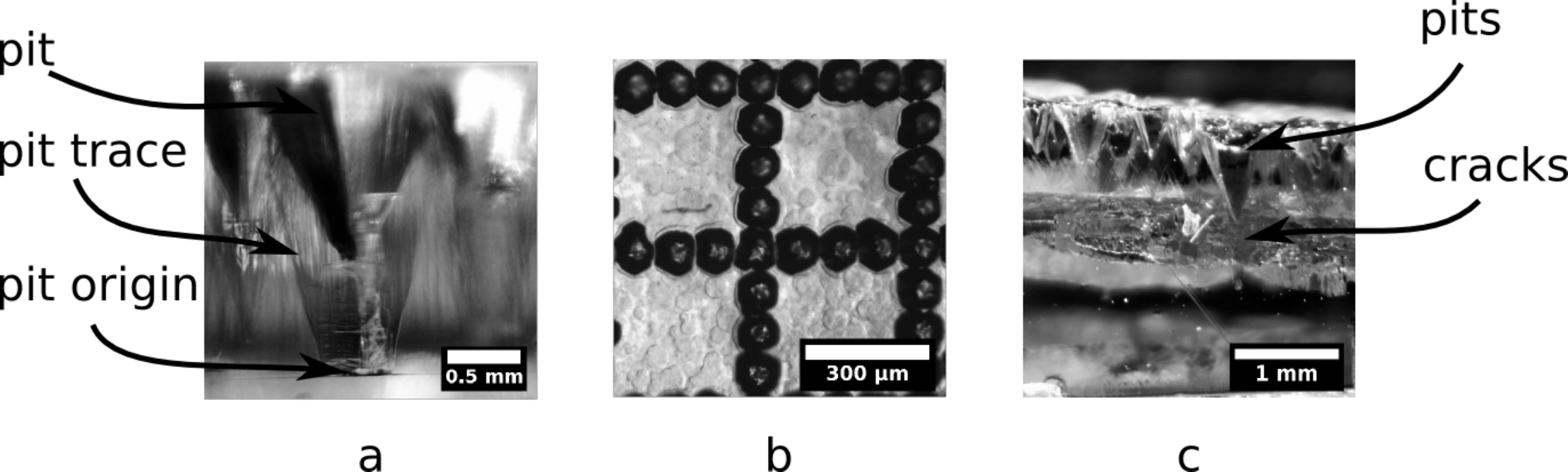}
  \caption[]{Typical origins of pit formation in HT mode:
  (a)~A pit formed over an area of foreign contamination (side view).
(b)~Pits formed over a network-shaped SiO$_2$ mask (top view).
(c)~Pits formed over cracks (side view).}
  \label{fig-originz}
  \end{figure}

The films with pits were investigated to determine the origin of pit formation.
The cross-sections of the films were prepared near the pits.
A study of these cross-sections showed that each pit leaves a visible trace during growth.
 An example of such a trace is shown in Fig.~\ref{fig-originz}(a).
 The material of the trace has a higher impurity concentration that affects  optical and electrical properties.
Such a trace can also be studied by photoluminescence, cathodoluminescence \cite{wagner2002influence},
 and selective etching~\cite{weyher2010revealing} measurements.
A study of pit traces allows us  to determine the history of pits from the moment of their formation.

\subsection{Origins of the pit formation}

It was found that pits were formed for various reasons.
The mechanism that prevailed depended on  growth parameters, the condition of the substrate surface, and the reactor design.

\subsubsection{  Substrate defects and contamination}
Pits develop on a film at early stages of its growth when the substrate contains defects.
A common source of such defects is bad polishing.
 The pits, in this case, decorate the scratches on the substrate surface.

A dielectric mask can be considered as an artificial defect of the substrate.
A film grown on a substrate with a SiO$_2$ mask is shown in  Fig.~\ref{fig-originz}(b).
The mask consisted of circles, arranged in a rectangular grid with dimensions of 400$\times$400~$\rm \mu m$.
The diameter of the circles was 50~$\rm \mu m$.
The distance between the centers of the circles was 100~$\rm \mu m$.
Pits were formed over each circle, repeating the mask pattern.

\subsubsection{Foreign particle}

A common source of foreign particles in an HVPE reactor is a polycrystalline  deposit in reactor parts.
The particles of this deposit detach from the place they were grown and fall on the growing crystal.
An example of a pit formed over a foreign particle is presented in Fig.~\ref{fig-originz}(a).
The highest pit density was observed under reactor parts with the heaviest parasitic deposition.
Optimization of gas flow in a reactor can greatly reduce parasitic deposition even at long growth processes and thus prevent foreign particle formation.

\subsubsection{Cracks}

 Cracks form in GaN layers during growth as a consequence of tensile growth stress \cite{etzkorn2001cracking,tsyuk2011effect}.
 Pits were often found to decorate the cracks, penetrating the surface of the GaN film.
The pits formed over a network of cracks are presented in Fig.~\ref{fig-originz}(c).
 
 Cracks can also serve as a source of gallium droplets, which in turn induce pit formation.
 GaN decomposition begins at crack faces during long growth processes.
 If a crack is not overgrown before a notable decomposition occurs,
  liquid gallium infiltrates the surface, where it reacts with ammonia and forms polycrystalline GaN.

 Crack formation could be inhibited by using the LT growth mode or the two-stage growth process.
 Another approach to prevent cracking was to grow spatially discontinuous layers, in which crystal geometry reduces the mechanical stress of any origin \cite{voronenkov2011patterned}.

\subsection{Overgrowth of pits}
 \begin{figure}
 \includegraphics[width=1\linewidth]{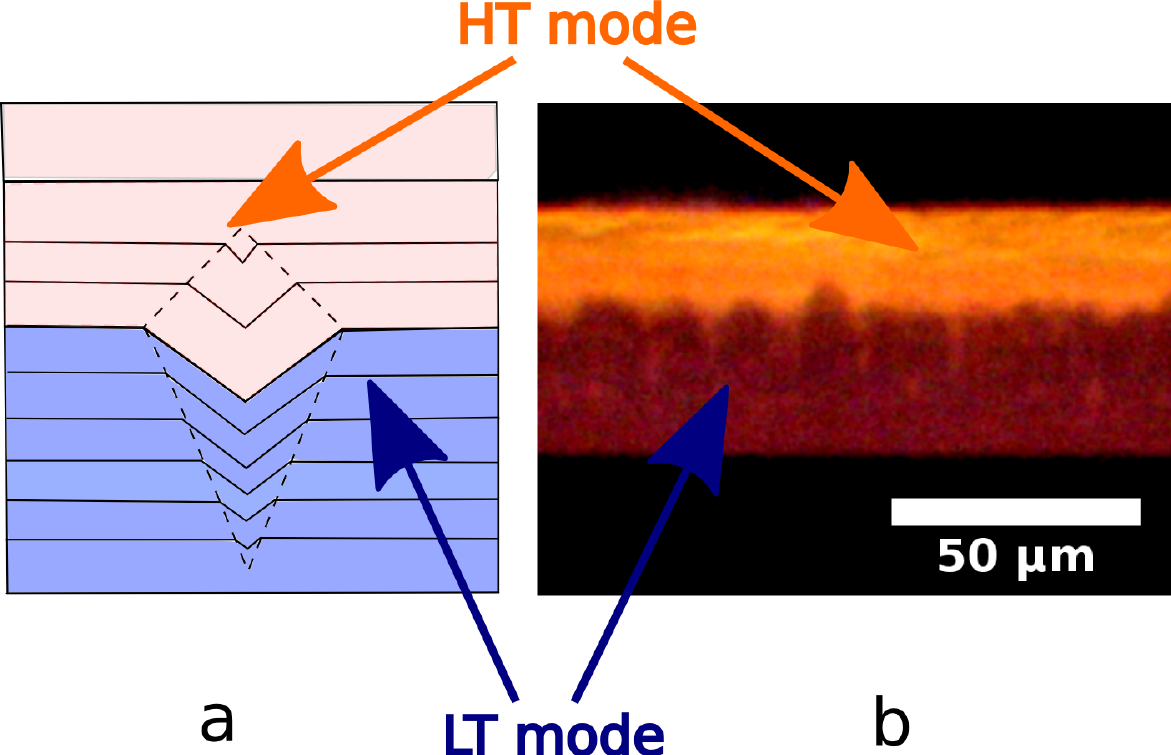}
 \caption[]{Pit overgrowth process caused by changes in growth parameters.
 (a)~A schematic of the pit overgrowth process caused by the transition from LT mode to HT mode.
(b)~A photoluminescence micrograph of a film grown in two stages.
The numerous pits formed during LT-mode growth were simultaneously overgrown when growth was switched to HT conditions.
 }
 \label{og-total}
 \end{figure}
A study of traces of overgrown pits, buried inside the films, allowed us to identify two mechanisms of overgrowth.

The first mechanism occurs during the two-stage process when the growth conditions are switched from the LT mode to the HT mode.
This results in an increase in the growth rate of pit facets, $v_{pit}$, and the overgrowth of pits formed during the first stage.
The pit shape remains constant during this process, and  growth parameters should be changed to alter the growth rate of pit facets.
The schematic of this process is presented in Fig.~\ref{og-total}(a).

 \begin{figure}
 \includegraphics[width=1\linewidth]{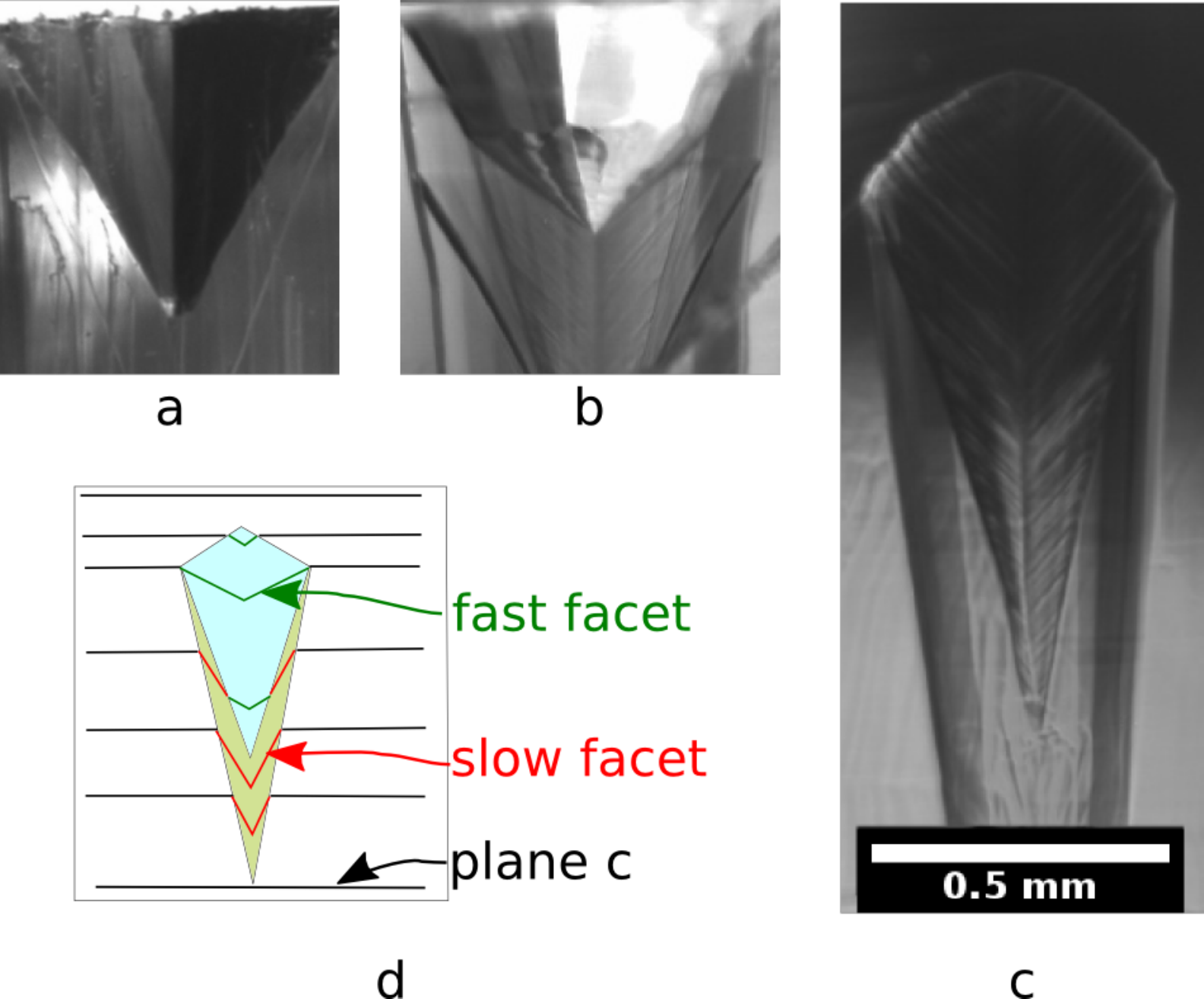}
 \caption[]{Stages of a spontaneous pit overgrowth process:
(a)~A pit formed by slow facets. This pit increased in size during growth.
(b)~A pit with  fast-growing facets developed at the bottom.
(c)~The whole pit was covered by fast facets and overgrown.
(d)~Schematic of spontaneous overgrowth process.
 }
 \label{og-some}
 \end{figure}
The second mechanism of overgrowth occurs spontaneously under constant growth conditions.
In contrast to the first mechanism,  overgrowth begins after a spontaneous change in pit shape, while growth rates of different facets remain unchanged.
The stages of this mechanism could be clearly seen on the cross sections of thick films containing pits (Fig.~\ref{og-some}).
The process of overgrowth begins with the nucleation of fast facets at the bottom of a pit.
The area occupied by these facets increases until they fill the pit and then the pit overgrows.

\section{Analysis}

   \begin{figure}
    \includegraphics[width=1\linewidth]{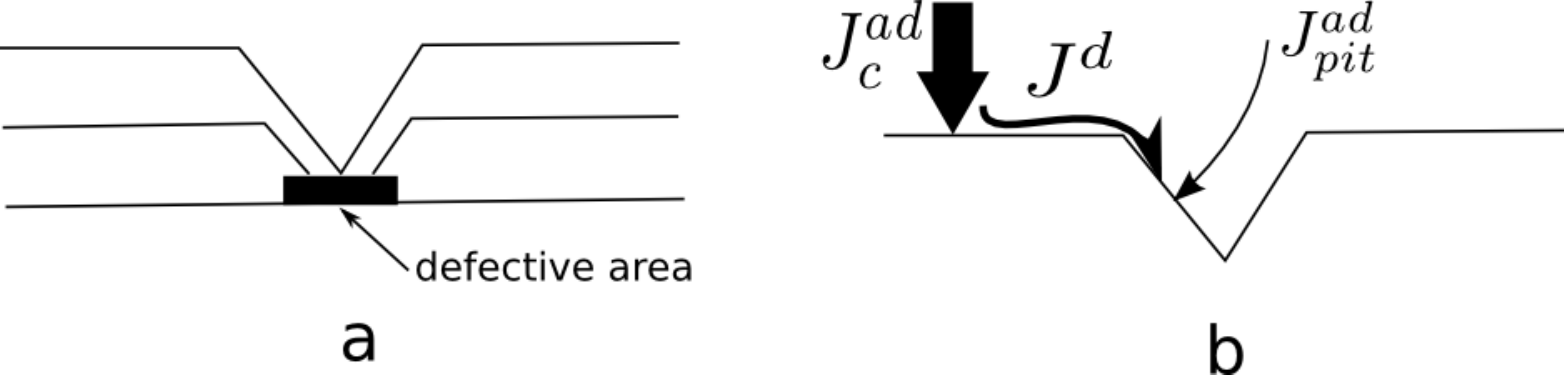}
   \caption[]{(a)~Schematic of the pit formation over the imperfection area with reduced growth rate.
   (b)~A pit facet with a sluggish adsorption kinetics adjoins with plane $\langle 0001 \rangle$ with rapid kinetics.
   The surface diffusion flux $J_d$ carries adsorbed species to the pit facet, increasing its growth rate.
  }
   \label{fig-pit-birth}
   \end{figure}

  \begin{figure}
   \includegraphics[width=1\linewidth]{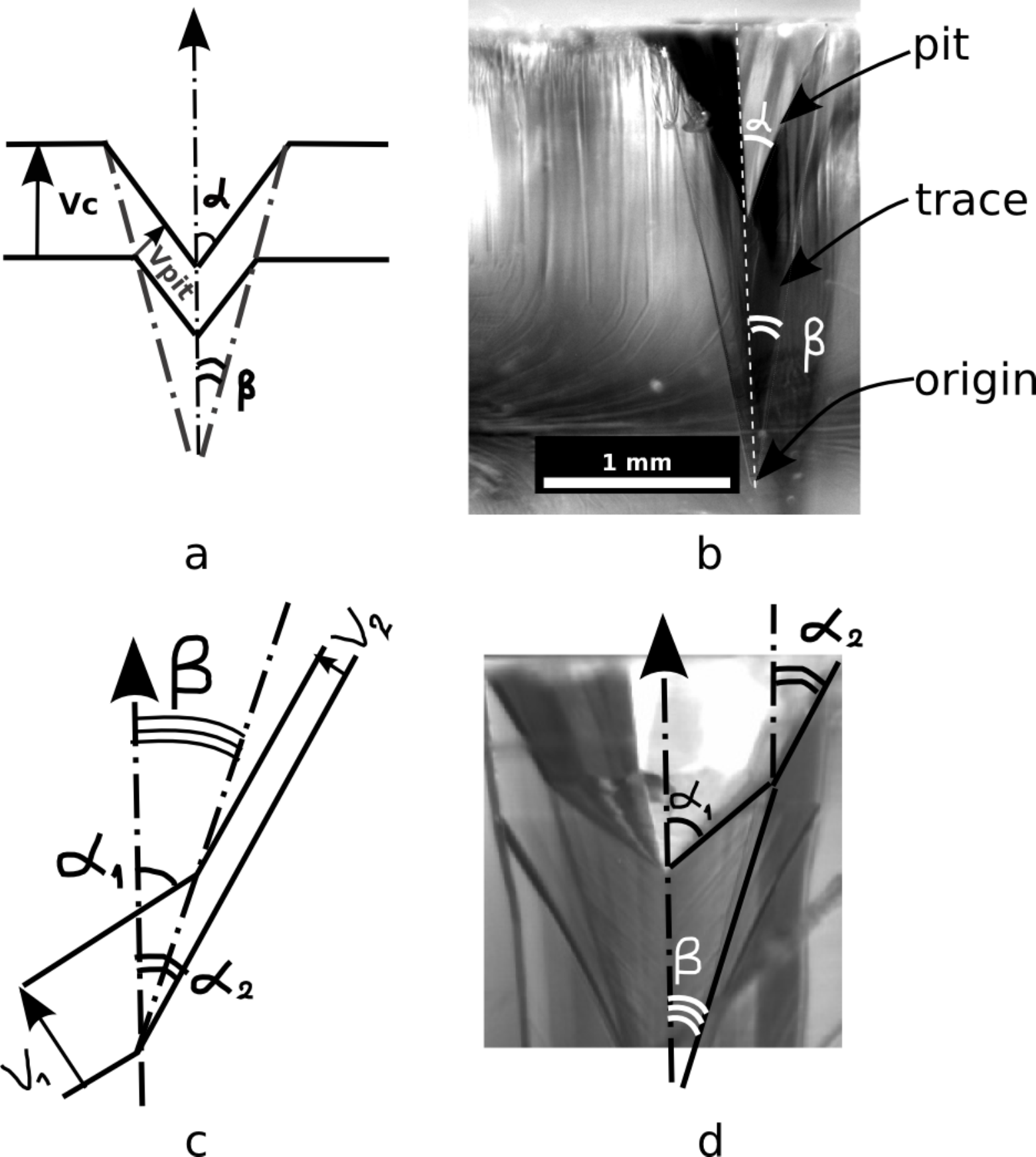}
  \caption[]{
 (a) Schematic of a pit on a plane $\langle 0001 \rangle$ at two consecutive moments of growth.
(b) Micrograph of a pit on a plane $\langle 0001 \rangle$. The shape of the pit and the pit trace can be seen.
(c) Schematic of two arbitrary adjacent facets at two consecutive moments of growth.
(d) Micrograph of a spontaneously overgrowing pit. Two adjacent  facets and the trace of their joint can be seen.
 }
  \label{fig-rates}
  \end{figure}
The origins of pit formation observed in the present work  can be generalized as
``imperfections that locally reduce growth rate''.
The origins of pit formation described in the literature, such as dislocations~\cite{chen1998pit} and inversion domains~\cite{lucznik2009bulk}, also fit this generalization.

  Pits also develop as a result of an incomplete coalescence of a spatially discontinuous film \cite{wagner2002influence}.
  In this case, the surface is initially jagged and pits form as a result of crystal habit evolution.

The schematic representation of pit formation  due to local growth rate reduction is presented in Fig.~\ref{fig-pit-birth}(a). 
The growth process proceeds normally everywhere except the defective area, which leads to a depression in the growth front.
By the time the defective area is completely overgrown, a faceted pit has formed.

Further evolution of the pit depends on the relationship between the growth rate of the $\langle 0001 \rangle$ plane, $v_c$, and the growth rate of facets composing the pit, $v_{pit}$.
If $v_{pit}$ is high as compared with $v_c$, the pit will likely overgrow, since fast facets tend to vanish as growth proceeds \cite{sunagawa}.

To obtain a quantitative relationship, consider the pit on the  $\langle 0001 \rangle$ plane at two consecutive moments of growth (Fig.~\ref{fig-rates}(a)).
During the period $\delta t = t_2-t_1$, the  $\langle 0001 \rangle$ plane advances to the distance  of $v_c \delta t$, while the pit plane advances to $v_{pit}\delta t$.
The angle of the pit trace, $\beta$, is considered to be positive if the pit size  increases with growth, and negative otherwise.
It can be shown that 
\begin{equation}
 \frac{v_{pit}}{v_{c}} = (\tan \alpha - \tan \beta)\cos \alpha .
\label{eq-rates}
\end{equation}
The pit will overgrow if 
\begin{equation}
 v_{pit} > v_{c}\sin\alpha .
\label{eq-criterion}
\end{equation}
If Eq. \ref{eq-criterion} holds for all possible pit planes, the plane  $\langle 0001 \rangle$ is morphologically stable.

The growth rate of a facet depends on two material fluxes: the flux of adsorption from the gas phase $J^{ad}$
and the surface diffusion flux from the adjacent facets $J^d$ (Fig.~\ref{fig-pit-birth}(b)).
If the size of the facet, $x$, is much greater than the average length of surface diffusion, $l_D$, the growth rate of the facet is governed solely by adsorption kinetics.
The adsorption kinetics, in turn, depends on the orientation of the facet, the density of steps on it, and other factors, such as contamination by impurities~\cite{parker1970growth}.

If  $x$ is less or comparable to  $ l_D$, the diffusion flux may be comparable to or exceed the condensation flux, thus altering the growth rate of the facet~\cite{volmer,wang2000mod}.
Surface diffusion tends to equalize the surface concentration of adsorbed species, thus increasing the growth rate of small facets with sluggish condensation kinetics.
Therefore, the pit will likely overgrow if the size of the defective region is smaller than  $ l_D$.

Kinematic considerations similar to Eq. \ref{eq-rates} can be applied to analyze  the process of spontaneous pit overgrowth (Figs. \ref{fig-rates}(c) and \ref{fig-rates}(d)).
This process occurs when a fast-growing facet with angle $\alpha_{1}$ appears at the bottom of the pit composed of planes with angle $\alpha_{2}$.
The area occupied by fast facets will increase or decrease according to the relationship between $v_{1}$ and $v_{2}$ and the angles of fast and slow facets ($\alpha_{1}$ and $\alpha_{2}$, respectively):

\begin{equation}
\tan\beta = \frac{v_2 \sin\alpha_1-v_1\sin\alpha_2}{v_2 \cos\alpha_1-v_1\cos\alpha_2} .
\end{equation}
Facet 1 will increase in size during growth if
\begin{equation}
   v_1\sin\alpha_2 > v_2\sin\alpha_1 .
\end{equation}
If at the same time 
$v_1 > v_c\sin\alpha_1$,
plane 1 will overrun plane 2 and fill the pit and
after that pit will overgrow.

\section{Conclusions}
Pits appear on a surface of a growing film owing to any defect that locally reduces growth rate.
The shape and density  of pits depend on the growth mode.
Pit density is high in the LT mode and can be reduced to nearly zero in the HT mode.
This can be explained by a higher surface diffusion in the HT mode.

Two mechanisms of pit overgrowth were observed.
Pits can be overgrown intentionally when  growth parameters are varied during growth so that the growth rate of pit facets increases and the criterion of overgrowth  (Eq. \ref{eq-criterion}) becomes valid.

A pit overgrows spontaneously if a fast-growing facet nucleates on its bottom.
This process occurs under constant growth conditions.

The formation of pits in the HT mode during long growth processes is induced by large-scale defects, such as foreign particles and cracks.
Careful tuning of the reactor design and growth recipe allows us to inhibit the formation of such defects and grow thick GaN films without pits.

\section*{Acknowledgment}
We gratefully acknowledge Uwe Popp and Marc Strafela of the University of Stuttgart for  FIB and AFM studies.

\pagebreak

\bibliographystyle{IEEEtran}
\bibliography{v-defects}

\end{document}